\documentclass[aps,showpacs,nofootinbib,superscriptaddress]{revtex4}
\usepackage{graphicx}
\usepackage{bm}
\usepackage{amsmath}
\usepackage{amssymb}
\usepackage{hyperref}

\newcommand{\qslash}{\kern 0.2 em n\kern -0.50em /}
\newcommand{\nslash}{\kern 0.2 em n\kern -0.50em /}
\newcommand{\kslash}{\kern 0.2 em k\kern -0.45em /}
\newcommand{\lslash}{\kern 0.2 em l\kern -0.50em /}
\newcommand{\pslash}{\kern 0.2 em p\kern -0.50em /}
\newcommand{\Sslash}{\kern 0.2 em S\kern -0.50em /}
\newcommand{\Pslash}{\kern 0.2 em P\kern -0.50em /}
\newcommand{\Dslash}{\kern 0.2 em D\kern -0.65em /\kern 0.15em}

\newcommand{\eps}{\epsilon}

\newcommand{\Tr}{\operatorname*{Tr}\nolimits}

\newcommand{\ii}{i}

\usepackage{color}

\begin{document}

\title{Predicting the $\sin\phi_S$ Single-Spin-Asymmetry of $\Lambda$ Production off transversely polarized nucleon in SIDIS}
\author{Yongliang Yang}
\email{yangyl@qdu.edu.cn}
\affiliation{College of Physics, Qingdao University, Qingdao 266071, China}
\author{Xiaoyu Wang}
\affiliation{School of Physics and Microelectronics, Zhengzhou University, Zhengzhou 450001, China}
\author{Zhun Lu}
\email{zhunlu@seu.edu.cn}
\affiliation{School of Physics, Southeast University, Nanjing 211189, China}

\begin{abstract}
We investigate the transverse target spin asymmetry $A^{\sin\phi_{S}}_{UT}$ for the unpolarized $\Lambda$ production in semi-inclusive deep inelastic scattering with the transverse momentum of the final-state lambda hyperon being integrated out.
The asymmetry is contributed by the product of the transversity distribution function $h_1(x)$ of the nucleon and the collinear twist-3 fragmentation function $\tilde{H}(z)$ of the $\Lambda$ hyperon.
The later one originates from the quark-gluon-quark correlation and is a naive time-reversal-odd function.
We calculate $\tilde{H}$ of the $\Lambda$ hyperon by adopt a diquark spectator model.
Using the numerical result of $\tilde{H}(z)$ and the available parametrization of $h_1(x)$ from SIDIS data, we predict the $\sin\phi_{S}$ asymmetry in the electroproduction of the $\Lambda$ hyperon in the kinematical region of EIC, EicC and COMPASS.
In the phenomenological analysis we include the evolution effect of the distribution functions and the fragmentation functions.
The results show that the asymmetries for the $\Lambda$ production SIDIS process is around 0.1 and may be accessible at EIC, EicC and COMPASS.
We also find that the evolution of fragmentation function can affect the size of asymmetry.
\end{abstract}

\maketitle

\section{introduction}

Understanding the nonperturbative fragmentation mechanism in hard semi-inclusive processes is one of the important tasks in hadronic physics.
Fruitful outcomes regarding fragmentation functions have been achieved by experimental measurements on $e^+ e^-$ annihilation~\cite{Abe:2005zx,Seidl:2008xc,TheBABAR:2013yha,Ablikim:2015pta,Guan:2018ckx}, semi-inclusive deeply inelastic scattering (SIDIS)~\cite{Airapetian:2004tw,Airapetian:2010ds,Qian:2011py,Adolph:2014zba} and $pp$ collision~\cite{Lesnik:1975my,Bunce:1976yb,Adams:2003fx,Abelev:2008af,Lee:2007zzh,Adamczyk:2012xd}, as well as by theoretical studies~\cite{Collins:1992kk,Boer:1997qn,deFlorian:1997zj,Boros:1998kc,Ma:1998pd, Ma:1999gj, Anselmino:2000vs,Bacchetta:2001di,Bacchetta:2002tk,Gamberg:2003eg,Bacchetta:2003xn,
Amrath:2005gv,Xu:2005ru,Bacchetta:2007wc,Zhang:2008ez,Anselmino:2013vqa,Kang:2014zza,Kang:2010xv,
Kanazawa:2013uia,Gamberg:2018fwy,Wei:2014pma}.
Of particular interests are those related to the spin-orbit correlations which are usually naive time-reversal-odd (T-odd).
A renowned fragmentation function is the Collins function $H_1^\perp$~\cite{Collins:1992kk},
which arises from the correlation between the transverse spin of the quark and the transverse momentum of the fragmented hadron.
Another example is the Sivers-type fragmentation function $D_{1T}^\perp$~\cite{Anselmino:2000vs,DAlesio:2020wjq,Callos:2020qtu} reflecting the correlation between the transverse polarization of the final state spin-1/2 hadron and the transverse momentum of the quark.
Because these functions describe the asymmetric distribution of hadron inside a fragmenting quark, they play important roles in the spin or azimuthal asymmetries in various high energy processes~\cite{Abe:2005zx,Seidl:2008xc,TheBABAR:2013yha,Ablikim:2015pta,Guan:2018ckx,Airapetian:2004tw,
Airapetian:2010ds,Qian:2011py,Adolph:2014zba,Lesnik:1975my,Bunce:1976yb,Adams:2003fx,Abelev:2008af,
Lee:2007zzh,Adamczyk:2012xd}.
Furthermore, these functions contain nontrivial QCD dynamics such as final-state interactions as well as the Wilson lines which ensure the gauge-invariance of the operator definitions~\cite{Metz:2002iz,Collins:2004nx, Boer:2003cm,Yuan:2007nd,Gamberg:2008yt,Meissner:2008yf}.

Recently, the twist-3 fragmentation functions arising from multiparton correlation~\cite{Qiu:1998ia,Eguchi:2006qz,Kang:2010zzb,Kanazawa:2014dca,Koike:2017fxr} also attract a lot of attentions.
Particularly, a phenomenological study~\cite{Kanazawa:2014tda} on the inclusive pion production in single transversely polarized $pp$ collision~\cite{Adams:2003fx,Abelev:2008af,Adamczyk:2012xd,Lee:2007zzh} shows that, beside the contribution from the twist three distribution $T_q(x,x)$~\cite{Efremov:1981sh,Efremov:1984ip,Qiu:1991wg,Kang:2011hk}, the T-odd twist-3 fragmentation functions $\hat H$ and $\tilde{H}$~\cite{Metz:2012ct} should be included in the analysis in order to interpret the single spin asymmetry (SSA) in this process in a consistent way.
The fragmentation function $\tilde{H}$ arising from quark-gluon-quark(qgq) correlation also contributes to the $\sin\phi_S$ asymmetry in SIDIS~\cite{Bacchetta:2006tn} through the combination $h_1(x) \otimes\tilde{H}(z)$, with $h_1(x)$ the tansversity distribution and $\phi_S$ the azimuthal angle of the transverse spin of the nucleon target.
Thus, $\tilde{H}(z)$ provides a unique manner to probe the transversity of the proton via single-hadron production in the collinear framework.
In Ref.~\cite{Lu:2015wja}, the fragmentation function of the pion meson was calculated by a quark-antiquark-spectator model.
In Ref.~\cite{Wang:2016tix}, the SSA $A_{UT}^{\sin(\phi_S)}$ for the pion production at the electron-ion collider (EIC) was predicted.

In this work, we will study the fragmentation function $\tilde{H}$ of the $\Lambda$ hyperon as well as its role in the SSA of $l p^\uparrow \to l^\prime \Lambda X $ process.
As the $\Lambda$ hyperon contains the up, down and strange valence flavors, which is more complicated than the pion meson, the study of the $\Lambda$ fragmentation function will provide complimentary information on hadronization mechanism involving spin-orbit correlation.
The investigation could also obtain the flavor dependence~\cite{Ma:1999gj} of the fragmentation process.
For this purpose, we calculate $\tilde{H}$ of the $\Lambda$ hyepron for the up, down, and strange quarks using a diquark spectator model.
Previously, the model has been applied to calculate the fragmentation function $D^\perp_{1T}$ and Collins functions of the $\Lambda$ hyperon in Refs.~\cite{Yang:2017cwi,Wang:2018wqo}.
In these cases the spectator system is diquark, and the contributions from both the scalar diquark and vector diquark are taken into account.
We also consider the gluon rescattering effect for the qgq correlation functions in the calculation.
Based on the model results, we predict the $\sin\phi_S$ asymmetry in the $l p^\uparrow \to l^\prime \Lambda X $ process, which can be measured by the COMPASS as well as the proposed EIC and the EIC in China (EicC). As these facilities cover different kinematical regions, it is necessary to consider the QCD evolution effect of $\tilde{H}$ to compare the result at different energy scales.

The remaining content of the paper is organized as follows.
in Sec.~\ref{Sec:2}, we perform the calculation for twist-3 qgq fragmentation function $\tilde{H}$ by adopting the diquark spectator model and  study the evolution effects of $\tilde{H}$.
In Sec.~\ref{sec:Aut}, we set up the formalism of the $\sin\phi_S$ asymmetry in SIDIS process with the transverse momentum of the final-state hadron being integrated.
We present the numerical results of the $\sin\phi_S$ asymmetry in the electroproduction of the $\Lambda$ hyperon at EIC, EicC and COMPASS, using formalism of the $\sin\phi_S$ asymmetry in a collinear framework.
In Sec.~\ref{conclusion}, we summarize the results of the paper and give some conclusion.

\section{Model calculation of fragmentation function in qgq correlator }
\label{Sec:2}

\begin{figure}
  \centering
  \includegraphics[width=6.5cm]{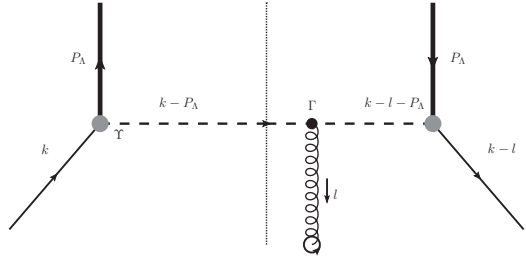}
  \caption{The Feynman diagram that is relevant to the calculation of the quark-gluon-quark correlator in the diquark model. The notation ${\Upsilon}$ and $\Gamma$ describe the quark-diquark-hyperon and gluon-diquark coupling vertex, respectively.}\label{fdqgq1}
\end{figure}

In this section, we present the model calculation of the twist-3 transverse momentum dependent fragmentation function $\tilde{H}(z,k_T)$ utilizing the diquark spectator model.
The fragmentation function can be obtained from the following trace of the transverse correlator $\tilde{\Delta}^\alpha_A(z,k_T)$
\begin{align}\label{trace}
{z\over 4M_\Lambda}\Tr[(\tilde{\Delta}_{A\alpha}(z,k_T;S_{\Lambda})+
\tilde{\Delta}_{A\alpha}(z,k_T;-S_{\Lambda}))\sigma^{\alpha-}]&=\tilde{H}(z,k_T)+\ii\tilde{E}(z,k_T)
\end{align}
where $M_\Lambda$ is the mass of $\Lambda$ hyperon, and the twist-3 qgq fragmentation correlator $\tilde{\Delta}_{A\alpha}$ can be expressed as~\cite{Gamberg:2006ru,Bacchetta:2006tn}
\begin{align}
\tilde{\Delta}_A^\alpha(z,k_T;S_{\Lambda}) &=\sum_{X}\hspace{-0.55cm}\int \; \frac{1} {2z}\int \frac{d\xi^{+}d^2\bm\xi_T} {(2\pi)^3}\int  e^{\ii k\cdot \xi} \langle 0| \int^{\xi^+}_{\pm\infty^+} d{\eta^+}\mathcal{U}^{\bm\xi_T}_{(\infty^+,\eta^+)}\nonumber\\
 &\times gF^{-\alpha}_\perp (\eta) \mathcal{U}^{\bm\xi_T}_{(\eta^+,\xi^+)} \psi(\xi)|P_{\Lambda},S_{\Lambda};X\rangle\langle P_{\Lambda},S_{\Lambda};X|\bar{\psi}(0)\mathcal{U}^{\bm 0_T}_{(0^+,\infty^+)}\mathcal{U}^{\infty^+}_{(\bm 0_T,\bm \xi_T)}|0\rangle\bigg|_{\begin{subarray}{l}
\eta^+ = \xi^+=0 \\ \eta_T = \xi_T \end{subarray}}\,.
\label{eq:qgq}
\end{align}
with ${\cal U}^{c}_{(a,b)}$ the Wilson line (gauge link) running along the direction from $a$ to $b$ at the fixed position $c$.
The detailed discussion on the Wilson line $\cal U$ has been given in Refs.~\cite{Bacchetta:2006tn,Bacchetta:2007wc,Ji:2002aa,Belitsky:2002sm}.
The gauge invariant of correlator is guaranteed by the antisymmetric field strength tensor $F^{\mu\nu}$ of the gluon.
The state $|P_\Lambda,S_\Lambda\rangle$ represents final-state lambda hyperon with the momentum of $P_\Lambda$ and the spin of $S_\Lambda$.
In the diquark spectator model~\cite{Nzar:1995wb,Jakob:1997wg}, the fragmentation function $\tilde{H}$ can be calculated from Fig.~\ref{fdqgq1}, where the contribution to the T-odd fragmentation function also originates from the imaginary part of the one-loop diagram~\cite{Yang:2016mxl,Lu:2015wja}.
Specifically, the quark-diquark-hyperon vertex $\langle\,P_\Lambda,S_\Lambda; X|\,\bar\psi(0)|0\rangle$ appearing in the r.h.s of Eq.(\ref{eq:qgq}) has the following form:
\begin{align}\label{eq:diquark}
  \langle\,P_\Lambda,S_\Lambda; X|\,\bar\psi(0)|0\rangle=
  \begin{cases}
  \bar{U}(P_\Lambda,S_\Lambda)\, {\Upsilon}_s\,\displaystyle{\frac{i}{\kslash-m_q}}&
  \textrm{scalar diquark,} \\
 \bar{U}(P_\Lambda,S_\Lambda\,){\Upsilon}^{\mu}_v \,\displaystyle{\frac{i}{\kslash-m_q}}\,\varepsilon_{\mu}\, & \textrm{axial-vector diquark.}
  \end{cases}
\end{align}
where $k$ denotes the parent quark momentum. $\varepsilon_{\mu}$ is the polarization vector of the axial-vector diquark.
The summation over all polarizations states of the axial-vector diquark can be expressed as $d_{\mu\nu}=\sum_\lambda\varepsilon^{*(\lambda)}_\mu\varepsilon^{(\lambda)}_\nu$, which has the form as $d_{\mu\nu}=-g_{\mu\nu}
+{\frac{P_{\Lambda\mu}\,P_{\Lambda\nu}}{M_\Lambda^2}}$~\cite{Jakob:1997wg}.
The scalar and axial-vector coupling vertex of the quark-diquark-hyperon can be expressed by ${\Upsilon}_s=g_s(k^2)$
and ${\Upsilon}^{\mu}_v ={g_v(k^2)\over\sqrt{3}}\gamma_5(\gamma^\mu+{P_\Lambda^\mu\over M_\Lambda})$, respectively.
In this work we assume that $g_s$ and $g_v$ have the same form $g_s=g_v=g_{qh}$ as Gaussian form denoted as $g_{qh}$:
 \begin{align}\label{factor}
  g_{qh}(k^2)~\mapsto {g_D\over z}\,e^{-{k^2\over \Lambda^2}}\,,
\end{align}
where $\Lambda^2$ has the general form $\Lambda^2= \lambda^2z^\alpha(1-z)^\beta$ and $g_D$, $\lambda$, $\alpha$ and $\beta$ are the model parameters.

Analogously, we also provide the expression for the vertex $\langle 0| gF^{-\alpha}_\perp (\eta)  \psi(\xi)|P_{\Lambda},S_{\Lambda};X\rangle$ in Eq.~(\ref{eq:qgq}), in which the Feynman rules corresponding to the gluon field strength tensor $F^{\alpha\beta}$ is given by the factor $\ii\,(l^\alpha g_T^{\rho\beta}-l_T^\alpha g^{\beta\rho})$ , as denoted by the open circle in Fig.~\ref{fdqgq1}~\cite{Lu:2015wja}.
With the above Feynman rules, the contribution of Fig.~\ref{fdqgq1} to the correlator $\tilde{\Delta}^\alpha_{A}$ is given by:
\begin{align}\label{eq:qgqFDsv}
\tilde{\Delta}^\alpha_{A\,s}(z,k_T,S_\Lambda)&=-i{C_F\alpha_S\over 2(2\pi)^2(1-z)P_\Lambda^-}{1\over k^2-m^2}\int{d^4 l\over (2\pi)^4}\notag\\
&\times{(l^-g_T^{\alpha\rho}-l^\alpha\,n_+^{\rho})(\kslash - \lslash + m)\bar{\Upsilon}_s U(P_\Lambda,S_\Lambda\,)\bar{U}(P_\Lambda,S_\Lambda\,)\Upsilon_s(\kslash+m)\over((k-l)^2-m^2)(l ^2-\ii\eps)((k-l-P_\Lambda)^2-m_s^2)(-l^--\ii\eps)}\bar{\Gamma}_{\rho}\\
\tilde{\Delta}^\alpha_{A\,v}(z,k_T,S_\Lambda)&=i{C_F\alpha_S\over 2(2\pi)^2(1-z)P_\Lambda^-}{1\over k^2-m^2}\int{d^4 l\over (2\pi)^4}\notag\\
&\times{(l^-g_T^{\alpha\rho}-l^\alpha\,n_+^{\rho})(\kslash - \lslash + m)\bar{\Upsilon}_v^\nu U(P_\Lambda,S_\Lambda\,)\bar{U}(P_\Lambda,S_\Lambda\,)\Upsilon_v^\mu(\kslash+m)\over((k-l)^2-m^2)(l ^2-\ii\eps)((k-l-P_\Lambda)^2-m_s^2)(-l^--\ii\eps)} d^{i\mu} d^{j\nu} \bar{\Gamma}_{i\,j\rho}
\end{align}
where $\tilde{\Delta}^\alpha_{A\,s}$ and $\tilde{\Delta}^\alpha_{A\,v}$ represent scalar and axial vector diquark form of qgq correlator, respectively.
Here, the light-cone coordinates $a^{\pm} = a\cdot n_{\pm} = (a^0\pm a^3)/\sqrt{2}$ is applied, and $k^{-}={P_\Lambda^{-}/{z}}$.
In Fig.~\ref{fdqgq1}, the notation $\Gamma$ describe the gluon-diquark coupling vertex.
In order to explicitly calculate the correlator, we choose the following form for the vertex between the gluon and the scalar diquark ($\Gamma_s$) and the axial vector diquark ($\Gamma_v$):
\begin{align}
\Gamma^{\rho,a}_s = &\ii\,T^a(2 k - 2 P - l)^\rho\,\\
\Gamma^{\rho\mu\nu,a}_v = &-\ii\,T^a[(2 k - 2 P - l)^ \rho\,g^{\mu\nu} -(k - P - l)^\mu g^{\nu\rho}-(k - P)^\nu g^{\rho\mu}];\label{eq:gcoupling}
\end{align}
Where $T^a$ is the Gell-Mann matrix.

Similar to the calculation of the T-odd quark-gluon-quark fragmentation function $\tilde{G}^\perp$ in Ref.~\cite{Yang:2016mxl}, we obtain the imaginary part of the correlator usig the Cutkosky cut rules to put the gluon and quark lines on the mass shell.
This corresponds to the following replacements on the propagators by using the Dirac delta functions
\begin{align}
{1\over l^2 + \ii\varepsilon} \rightarrow -2\pi i\delta(l^2),~~~~~~~ {1\over (k-l)^2-m^2 + \ii\varepsilon} \rightarrow -2\pi i\delta((k-l)^2-m^2) \,.\label{eq:cuts}
\end{align}

Using the cut rules in Eq.~(\ref{eq:cuts}), we perform the trace and integration over the loop momentum $l$, we first give the scalar diquark component of $\tilde{H}(z,k_T^2)$ for $\Lambda$ hyperon as follows
\begin{align}
\tilde{H}^{s}(z,\bm{k}_T^2)={\alpha_S\,g_{qh}^2\,C_F \over \,(2\pi)^4M_\Lambda(1-z)}{1\over k^2-m^2}\tilde{H}_{1s}(z,\bm{k}_T^2)
\end{align}
where
\begin{align}
 \tilde{H}_{1s}(z,\bm k_T^2)&={1\over z}[z\mathcal{A}(z^2\bm{k}^2_T m +k^2 (z-1)(zm+M_\Lambda))+\mathcal{B} M_\Lambda(z(z-1)m M_\Lambda -z^2\bm{k}^2_T +(z-1)M_\Lambda^2)].
\end{align}
Here $k^2 = z \bm{k_T}^2/(1-z) + m_s^2/(1-z) + m_h^2/z$.

Similarly, using the gluon-diquark vertex given in Eq.~(\ref{eq:gcoupling}), we
can also calculate the expression for $\tilde{H}$ from the axial
vector diquark component
\begin{align}\label{eq:vresult}
\tilde{H}^v(z,k_T^2)={ \alpha_S\,g_{qh}^2\,C_F \over 4(2\pi)^4M_\Lambda(1-z)}{1\over k^2-m^2}(\tilde{H}^v_{s}(z,\bm{k}_T^2)+\tilde{H}^v_{0}(z,\bm{k}_T^2)+
\tilde{H}^v_{1}(z,\bm{k}_T^2)+\tilde{H}^v_{2}(z,\bm{k}_T^2))
\end{align}
where the four terms in the r.h.s. of Eq. (\ref{eq:vresult}) are given by
\begin{align}
\tilde{H}^v_{s}&={4\over z}[(z\mathcal{A}(z^2\bm{k}^2_T m+k^2 (z-1)(zm+M_\Lambda))+\mathcal{B} M_\Lambda(z(z-1)m M_\Lambda -z^2\bm{k}^2_T +(z-1)M_\Lambda^2)]\,,\\
\tilde{H}_{0}&=-2z {\bm k_T^2\over 3 M_\Lambda} [(2 \mathcal{CC}k^-k\cdot P+2 \mathcal{CD}k^- M_\Lambda^2+2 z \mathcal{CE}k^-k^- -\mathcal{C}k^- k^2 z+\mathcal{C}k^-m^2 z)]\,,\\
\tilde{H}_{1}&={2\over 3 M_\Lambda^2 z} \bigg{\{}z k\cdot P [z (\bm{k}^2_T(\mathcal{AA} m z-2 \mathcal{A} (m z+M_\Lambda))+\mathcal{AA} k^2 (m z+M_\Lambda)\notag\\
&-2\mathcal{A} k^2 m z-5\mathcal{A} k^2 M_\Lambda-3 \mathcal{A} m^2 M_\Lambda-2 \mathcal{B} m M_\Lambda^2)+\mathcal{AB} M_\Lambda^2 (m z-3 M_\Lambda)]\notag\\
&-2 m_h z (\mathcal{AA}-2\mathcal{A}) (k\cdot P)^2+M_\Lambda[z (M_\Lambda (\mathcal{AB} k^2 m z^2+\mathcal{AB}k^2 M_\Lambda z+\mathcal{BB} M_\Lambda^2 (m z-M_\Lambda)\notag\\
&-2 \bm{I}_2 (k^2-m^2) (m z+M_\Lambda)+m \mathcal{W}_1 z-M_\Lambda \mathcal{W}_1)+\bm{k}^2_T(\mathcal{AB} m M_\Lambda z^2+\mathcal{A} (z^2 (k^2+m^2)-4 M_\Lambda^2))\notag\\
&+\mathcal{A} k^2 (z^2 (k^2+m^2)+4 m M_\Lambda z+4 M_\Lambda^2))+2\mathcal{B} M_\Lambda^2 (-z^2 (\bm{k}^2_T+m^2)+2 m M_\Lambda z+2 M_\Lambda^2)]\bigg{\}}\,,\\
\tilde{H}_{2}&={4\over 3M_\Lambda}[M_\Lambda (\mathcal{A}+\mathcal{B} z)(M_\Lambda (2 \bm{k}^2_T+k^2+m^2)+2 m k\cdot P)-(k\cdot P+m M_\Lambda)((\mathcal{AA}+\mathcal{AB}z) k\cdot P\notag\\
 &+(\mathcal{AB}+z\mathcal{BB})M_\Lambda^2+z\mathcal{W}_1))]\,.
\end{align}
Here the number of the subscript for $\tilde{H}$ denote the number of the factor $l^-$ in the numerator of Eq.~(\ref{trace}) after the trace calculation is performed.
$\mathcal{A}$, $\mathcal{B}$, $\mathcal{C}$ and $\bm{I}_2$ are functions of $k^2$, $m$, $m_D$ and $M_\Lambda$, and can be found in Ref.~\cite{Yang:2017cwi}.
The functions $\mathcal{AA}$, $\mathcal{AB}$, $\mathcal{BB}$, $\mathcal{CC}$, $\mathcal{CD}$, $\mathcal{CE}$ and $\mathcal{W}_1$ are come from the following double-$l$ integrals
\begin{align}
\int{d^4l}&{l^\mu\,l^\nu\,\delta(l^2)\delta((k-l)^2-m^2)\over ((k-P_h-l)^2-m_s^2)}=\mathcal{AA}k^\mu\,k^\nu+\mathcal{BB}P^\mu\,P^\nu+
\mathcal{AB}(k^\mu\,P^\nu+P^\mu\,k^\nu)+g^{\mu\nu}\mathcal{W}_1\,,\label{eq:l1}\\
\int{d^4l}&{l^\mu\,l^\nu\,\delta(l^2)\delta((k-l)^2-m^2)\over ((k-P_h-l)^2-m_s^2)(-l\cdot n_++i\epsilon)}=\mathcal{CC}_fk^\mu\,k^\nu+\mathcal{DD}_fP^\mu\,P^\nu+\mathcal{EE}_fn_+^\mu\,n_+^\nu\notag\\
&+\mathcal{CD}_f(k^\mu\,P^\nu+P^\mu\,k^\nu)+\mathcal{CE}_f(k^\mu\,n_+^\nu+n_+^\mu\,k^\nu)
+\mathcal{DE}_f(n_+^\mu\,P^\nu+P^\mu\,n_+^\nu)+g^{\mu\nu}\mathcal{W}_2\,. \label{eq:l2}
\end{align}
Note that a main difference between the calculation for the $\Lambda$ hyperon and the one for the pion~\cite{Lu:2015wja} is that there are double $l$ in the numerators in Eqs.~(\ref{eq:l1}) and (\ref{eq:l2}), which need to be evaluated carefully.
We provide the complete expressions for these functions in the Appendix.

\begin{figure}
\centering
\includegraphics[width=7.5cm]{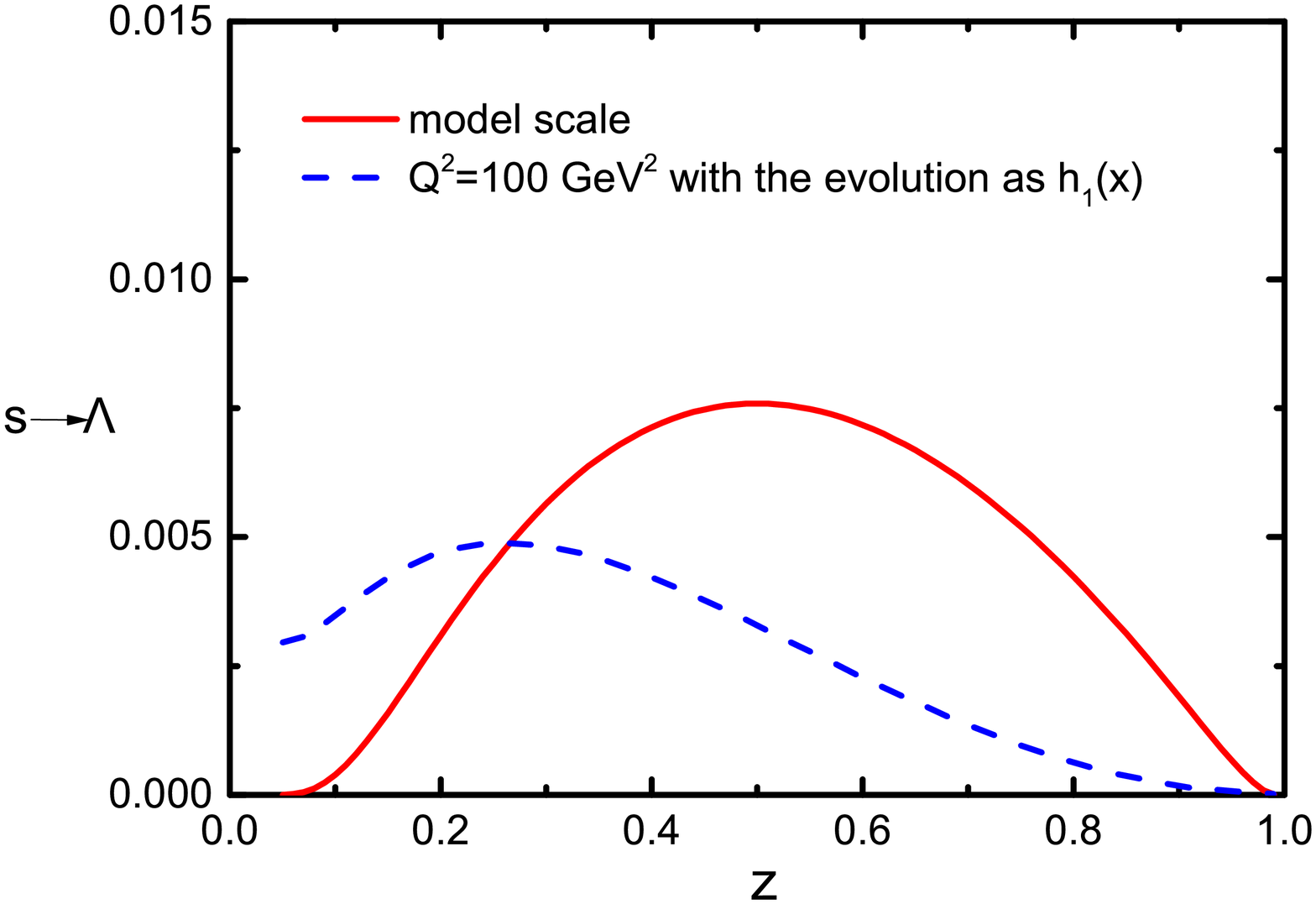}
\includegraphics[width=7.5cm]{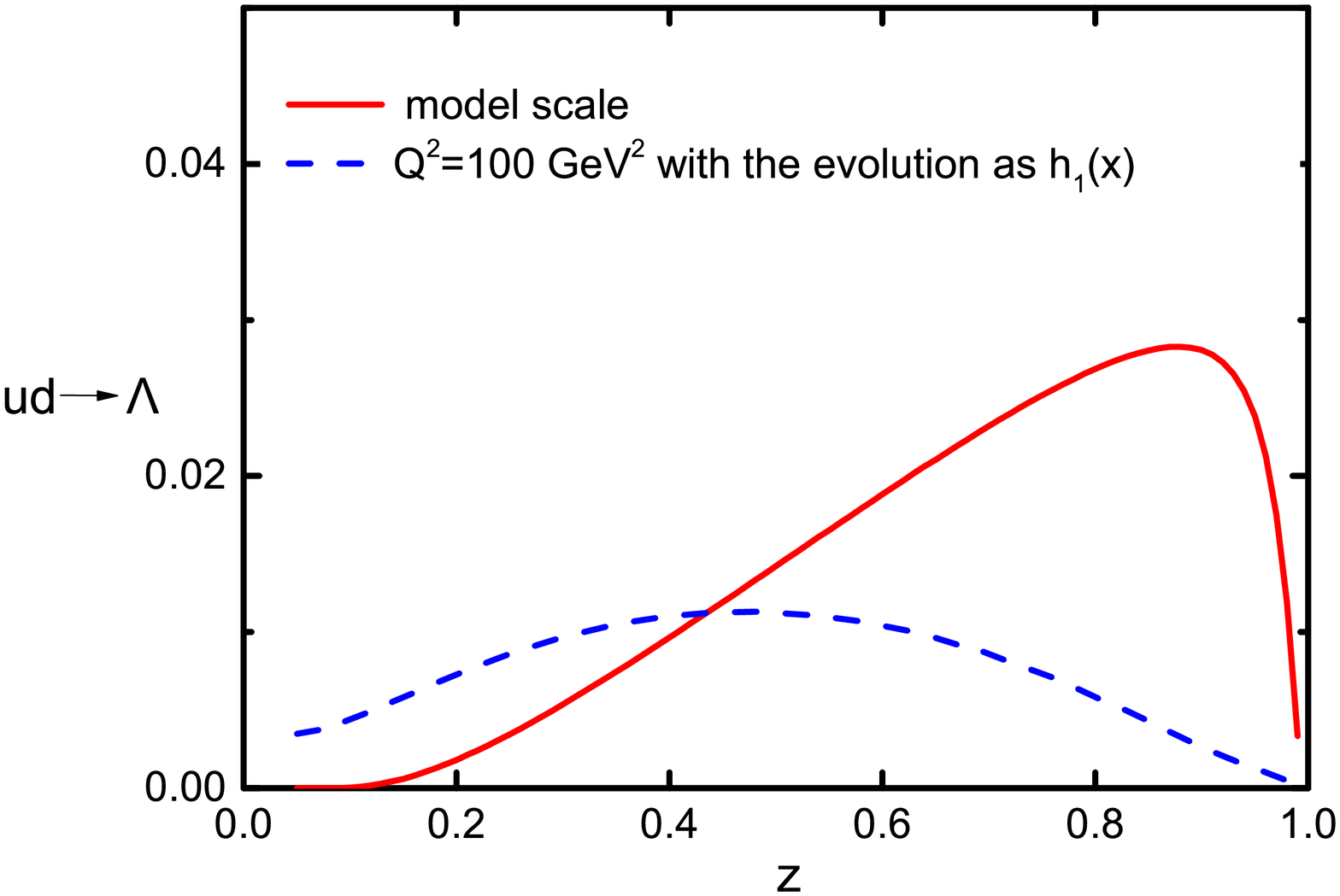}
\caption{Result of $z\tilde H^{\Lambda/s}(z)$ (left panel) and  $z\tilde H^{\Lambda/u(d)}(z)$ (right panel) at the model scale $Q^2\ =0.23\,\mathrm{GeV}^2$ (red solid lines) and the evolved results at $Q^2=100\,\mathrm{GeV}^2$ (blue dashed lines).}
\label{fig:htilde}
\end{figure}

We will focus on the favored quark contribution to the fragmentation function of $\Lambda$, i.e., $u\rightarrow\Lambda$ or $s\rightarrow \Lambda$, while the unfavored quark contribution is zero.
Assuming an SU(6) spin-flavor symmetry, the fragmentation functions of the $\Lambda$ hyperon for light flavors satisfy the following relations between different quark flavors
and diquark types~\cite{VanRoyen:1967nq,Jakob:1993th}
 \begin{align}\label{relation}
D^{\textrm{u}\rightarrow \Lambda} =\,D^{\textrm{d}\rightarrow \Lambda} ={1\over 4}D^{(s)}+{3\over 4}D^{(v)}\,,~~D^{\textrm{s}\rightarrow \Lambda}=D^{(s)}\,,
\end{align}
where $u$, $d$ and $s$ denote the up, down and strange quarks, respectively.
In this study we assume that the relation in Eq.~(\ref{relation}) holds for all fragmentation functions.
Neglecting the mass differences between the up, down and strange quarks ($m=0.36$ GeV), we obtain the light flavors fragmentation function $\tilde{H}$ as follows
\begin{align}\label{relationH}
\tilde{H}^{\textrm{u}\rightarrow \Lambda} =\tilde{H}^{\textrm{d}\rightarrow \Lambda} ={1\over 4}\tilde{H}^{(s)}+{3\over 4}\tilde{H}^{(v)}\,,~~\tilde{H}^{\textrm{s}\rightarrow \Lambda}=\tilde{H}^{(s)}\,,
\end{align}
In order to obtain the numerical result for $\tilde{H}$, we should choose the values for the parameters as Ref.~\cite{Yang:2017cwi}
\begin{align}\label{paramters}
g_D=1.983,~~~ m_D=0.745\textrm{GeV},~~~ \lambda=5.967\textrm{GeV},~~~ \alpha=0.5(\textrm{fixed}),~~~ \beta=0(\textrm{fixed})\,,
\end{align}
the parameters were obtained by fitting the model result of $D_1^\Lambda(z)$ in the same model to the deFlorian-Stratmann-Vogelsang (DSV) parametrization \cite{deFlorian:1997zj} at the model scale $Q^2=0.23\,\text{GeV}^2$.

Using the model parameters, we calculate the collinear twist-3 fragmentation function $\tilde{H}(z)$ of $\Lambda$ by integrating over the transverse momentum:
\begin{align}
\tilde{H}(z)=z^2\int d^2\bm{k_T}\tilde{H}(z,\bm{k_T}^2).
\end{align}
Since the energy scale in the experiments covers a wide range of $Q$, which is much higher than the model scale, it is necessary to include the QCD evolution effects of the fragmentation functions.
There are studies on the evolution of several twist-3 fragmentation functions in the literature~\cite{Kang:2010xv,Kang:2010zzb,Kang:2014zza,Ma:2017upj}.
However, the DGLAP evolution kernel for $\tilde{H}$ is still unknown.
In this work, we assume the evolution kernel of $\tilde{H}$ is same as the homogenous terms in the kernel of $\hat{H}^{(3)}(z)$ in Ref.~\cite{Kang:2010xv}, as has been done in Ref.~\cite{Kang:2015msa}.
This kernel has the same form as that of transversity distribution function.
In Ref.~\cite{Wang:2016tix}, the evolution kernel for the pion $\tilde{H}(z)$ has also been adopted as same as that for the transversity distribution function $h_1$, which was motivated by the fact that $\tilde{H}$ is also a chiral odd fragmentation function.
To do this, we apply the QCDNUM package~\cite{Botje:2010ay} and customize the package to include the kernel of transversity to perform the evolution.

In the left and right panels of Fig.~\ref{fig:htilde}, we plot the $z-$dependence of the collinear twist-3 fragmentation function $z\tilde{H}(z)$ of $\Lambda$ for $s$ and $u(d)$ quark of the twist-3 fragmentation functions at the model scale $Q^2 = 0.23 \mathrm{GeV}^2$ (the solid lines) and the evolved results at $Q^2 = 100 \mathrm{GeV}^2$ (the dotted lines), respectively.
Fig.~\ref{fig:htilde} shows that the magnitude of $z\tilde{H}(z)$ for $u(d)\rightarrow \Lambda$ increases with increasing $z$ when $0<z<0.9$, while it decreases with increasing $z$ after $z>0.9$ at the model scale, while the peak is around $z=0.5$ for the $s$ quark.
The evolved fragmentation functions at $Q^2=100\mathrm{GeV}^2$ show the relatively strong impact of the evolution effects.
The $z$-dependences of the $u(d)$ and $s$ quark for $z\tilde{H}$ is obvious different in the entire $z$ region.
We can see that the evolution from low $Q$ to higher $Q$ increases the sizes of $u(d)\rightarrow \Lambda$ in the region $z<0.4$, while the region for strange quark is $z<0.25$.
This is because $s$ quark only come from the contribution of the scalar diquark component in Eq.~(\ref{relationH}).

\section{Prediction on the $\sin \phi_S$ transverse SSAs of $\Lambda$ hyperon production in SIDIS}

\label{sec:Aut}
\begin{figure}
  \centering
  \includegraphics[width=7cm]{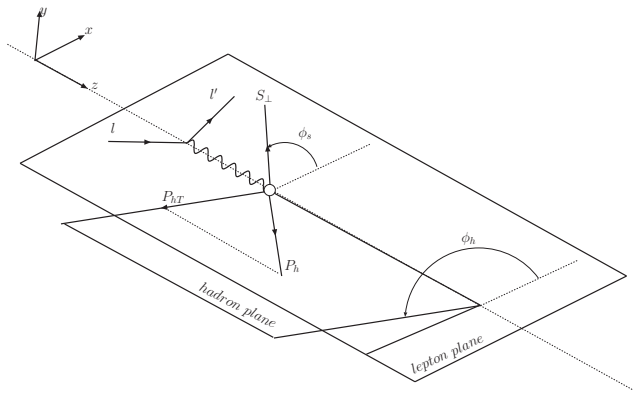}
  \caption{The definition of the azimuthal angles in SIDIS. $P_{h}$ stands for the momentum of the $\Lambda$ hyperon hadron, $S_\perp$ is the transverse component of the spin vector $S$ with respect to the virtual photon momentum~\cite{Bacchetta:2006tn}.\label{sidisplane}}
\end{figure}

The process under study is the unpolarized $\Lambda$ production in SIDIS process with an unpolarized lepton beam colliding on a transversely polarized nucleon beam (or target):
\begin{equation}
\label{eq:sidis}
l(\ell)+N^\uparrow(P) \longrightarrow l(\ell^\prime)+{\Lambda}(P_\Lambda)+X(P_X),
\end{equation}
where $l$ and $l'$ stand for the momenta of the incoming and outgoing leptons, $P$ and $P_\Lambda$ denote the momenta of the target nucleon and the final-state $\Lambda$ hyperon, respectively.
The momentum of the exchanged virtual photon is defined as $q=l-l'$ and $Q^2=-q^2$.
The reference frame in this work are adopted as Fig.~\ref{sidisplane}, in which we will consider the case the polarization of detected $\Lambda$ hyperon is not measured. The azimuthal angle $\phi_S$ stands for the angle between the lepton scattering plane and the direction of the transverse spin of the nucleon target.

We introduce the invariant variables to express the differential cross section as follows~\cite{Bacchetta:2006tn}
\begin{align}
x&=\frac {Q^2}{2P \cdot q}\,,\quad y=\frac{P \cdot q}{P \cdot l}\,,\quad z=\frac{P \cdot P_h}{P \cdot q}\,,\nonumber\\
 \gamma&=\frac{2Mx}{Q}\,,\quad W^2=(P+q)^2\,,\quad s=(P+l)^2\,. \nonumber
\end{align}
where $M$ denotes the mass of the target nucleon.
With the invariant variables, the differential cross section of the process for unpolarized $\Lambda$ production in SIDIS off an transverse polarized target can be expressed as ~\cite{Wang:2016tix,Bacchetta:2006tn}
\begin{align}
&\frac{d^6\sigma}{dxdydzd\phi_h d\phi_S dP_{hT}^2}=\frac{\alpha^2}{xyQ^2}\frac{y^2}{2(1-\varepsilon)}(1+\frac{\gamma^2}{2x})\nonumber\\
&\times \left\{F_{UU,L}(x,z,P_T)+ |S_\perp|[\sin \phi_S \,F^{\sin\phi_S}_{UT}(x,z,P_T)\right.\nonumber\\
&+\sin (2\phi_h-\phi_S) \,F^{\sin(2\phi_h-\phi_S)}_{UT}(x,z,P_T) \nonumber\\
& + \left. \textrm{leading twist terms}]\right\}\,, \label{eq:diffcs1}
\end{align}
where $\varepsilon$ is the ratio of the longitudinal and transverse photon flux
\begin{equation}
\label{eq:epsilon}
\varepsilon=\frac{1-y-\frac{1}{4}\gamma^2y^2}{1-y+\frac{1}{2}y^2+\frac{1}{4}\gamma^2y^2}.
\end{equation}
After integrating over $\bm P_{hT}$, the differential cross section in Eq.~(\ref{eq:diffcs1}) turns to the form
\begin{align}
&\frac{d^4\sigma}{dxdydzd\phi_S}=\frac{2\alpha^2}{xyQ^2}\,\frac{y^2}{2(1-\varepsilon)}\,
\left(1+\frac{\gamma^2}{2x}\right)\nonumber\\
&\times \,\left\{F_{UU,L}(x,z)+|S_\perp|\,\sqrt{2\varepsilon(1+\varepsilon)}\sin \phi_S\, F^{\sin\phi_S}_{UT}\left(x,z\right)+\cdots\right\}\,. \label{eq:diffcs2}
\end{align}
Here, the non-vanishing integrated structure functions are as follows~\cite{Boer:1997nt,Bacchetta:2006tn}
\begin{align}
F_{UU}(x,z)&=x\sum_q e_q^2f_1^q(x)D_1^q(z)\,,\label{eq:futcol1}\\
F^{\sin\phi_S}_{UT}\left(x,z\right) & =  -x\sum_q e_q^2\frac{2M_\Lambda}{Q}h_1^q(x)\frac{\tilde H^q(z)}{z}\,,\label{eq:futcol2}
\end{align}
and only the convolution of the transversity and the twist-3 collinear fragmentation function $\tilde{H}$ remains in structure function $F^{\sin\phi_S}_{UT}$.
Following Eqs.~(\ref{eq:futcol1}) and (\ref{eq:futcol2}), we can obtain the $z$ dependent-$\sin \phi_S$ asymmetry:
\begin{align}
\label{eq:autz}
&A_{UT}^{\sin\phi_s}(z) \nonumber\\
&=\frac{\int dx\int dy\frac{\alpha^2}{xyQ^2}\frac{y^2}{2(1-\epsilon)}(1+\frac{\gamma^2}{2x})
\sqrt{2\epsilon(1+\epsilon)}F^{\sin\phi_s}_{UT}(x,z)}
{\int dx\int dy\frac{\alpha^2}{xyQ^2}\frac{y^2}{2(1-\epsilon)}(1+\frac{\gamma^2}{2x})F_{UU}(x,z)}\,.
\end{align}
In a similar way, $x$ dependent $\sin \phi_S$ asymmetry can be written as
\begin{align}
\label{eq:autx}
&A_{UT}^{\sin\phi_S}(x) \nonumber \\
=&\frac{\int dy\int dz\frac{\alpha^2}{xyQ^2}\frac{y^2}{2(1-\epsilon)}(1+\frac{\gamma^2}{2x})\sqrt{2\epsilon(1+\epsilon)}F^{\sin\phi_s}_{UT}(x,z)}
{\int dy\int dz\frac{\alpha^2}{xyQ^2}\frac{y^2}{2(1-\epsilon)}(1+\frac{\gamma^2}{2x})F_{UU}(x,z)}\,,
\end{align}

In order to estimate the $\sin \phi_S$ asymmetry, we need the information of the transversity distribution~$h_1^q(x)$, for which we adopt the parametrization from Ref.~\cite{Anselmino:2013vqa}:
\begin{equation}
\label{eq:transversity}
h_1^q(x)=\frac{1}{2}\mathcal{N}_q^T(x)[f_{1}^q(x)+g_1^q(x)]\,,
\end{equation}
with
\begin{equation}
\label{eq:n}
 \mathcal{N}_q^T(x)=N_q^T\,x^\alpha(1-\beta)^\beta\frac{(\alpha+\beta)^{\alpha+\beta}}{\alpha^\alpha\beta^\beta}\,.
\end{equation}
where values of the parameters $N_u^T=0.36,\, N_d^T=-1.00,\,  \alpha=1.06$, and $\beta=3.66$ are taken from Ref.~\cite{Anselmino:2013vqa}.
The parametrization for the unpolarized distribution $f_1^q(x)$ is from Ref.~\cite{Lai:2010vv} and that for the helicity distribution $g_1^q(x)$ is from Ref.~\cite{deFlorian:2008mr}, respectively.
We note that currently there is no available parametrization on $h_1^q(x)$ for the sea quarks, therefore, in this calculation we will not consider the contribution from the transversity of the sea quarks.

The kinematical region of the EIC adopted in our calculation is~\cite{Accardi:2012qut}
\begin{align}
& 0.001<x<0.4,\quad 0.01<y<0.95,\quad 0.2<z<0.8,\nonumber\\
& Q^2>1 \mathrm{GeV}^2, \quad \sqrt{s}=45\ \mathrm{GeV},\quad W>5\ \mathrm{GeV}.
\end{align}
where $W$ is invariant mass of the virtual photon-nucleon system and $W^2\approx \frac{1-x}{x}Q^2$.
As for the EicC, we adopt the following kinematical cuts~\cite{eicc,Anderle:2021wcy}
\begin{align}
&0.005<x<0.5,\quad 0.07<y<0.9, \quad 0.2<z<0.7,\nonumber\\
&Q^2>1 \mathrm{GeV}^2,\quad \sqrt{s}=16.7\ \mathrm{GeV},\quad W>2 \mathrm{GeV}.
\end{align}
As the kinematics at EIC and EicC cover a wide range of $Q$, the QCD evolution effect of the distribution and fragmentation functions are also considered.

\begin{figure}
  \centering
  \includegraphics[width=7.5cm]{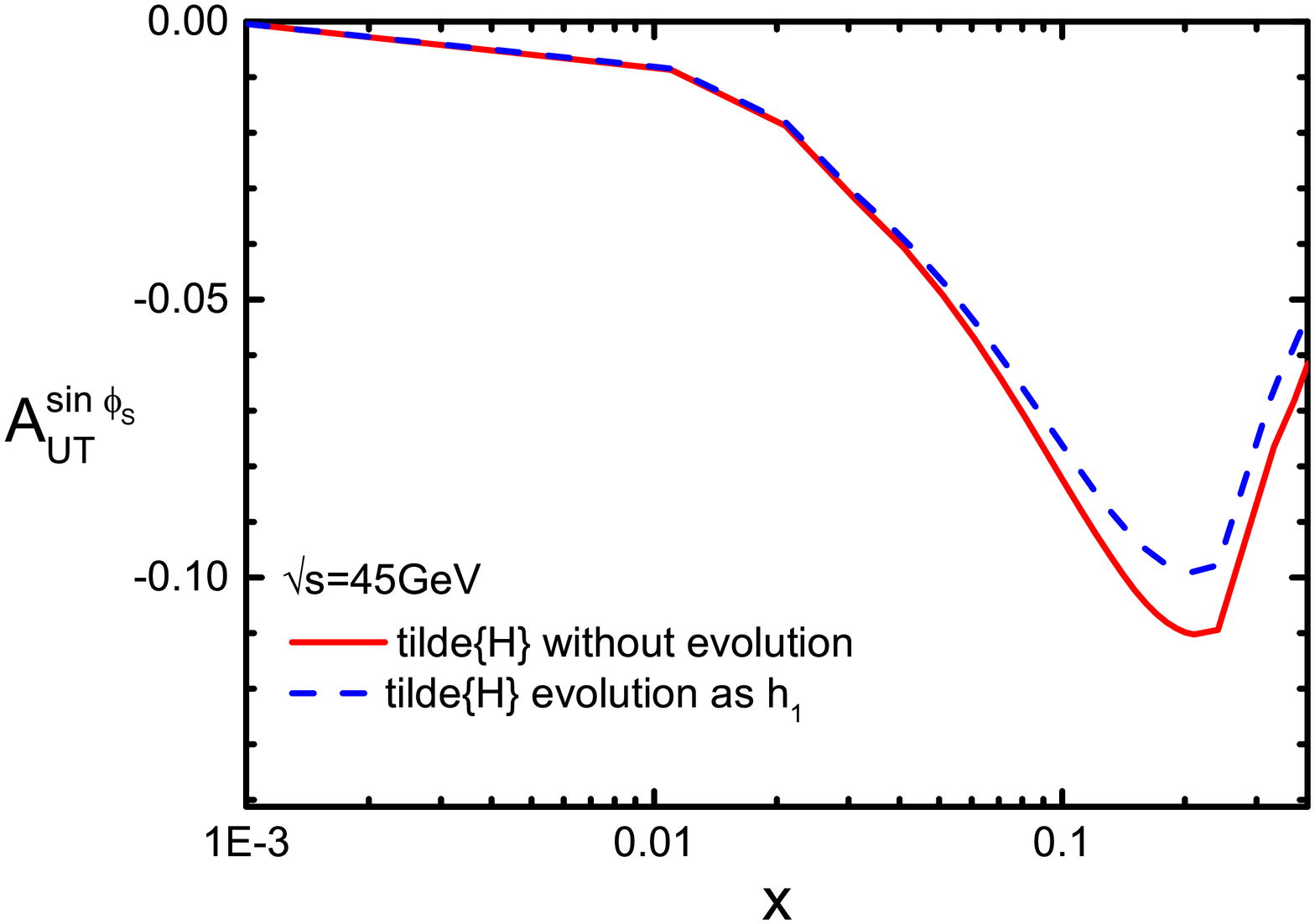}
  \includegraphics[width=7.5cm]{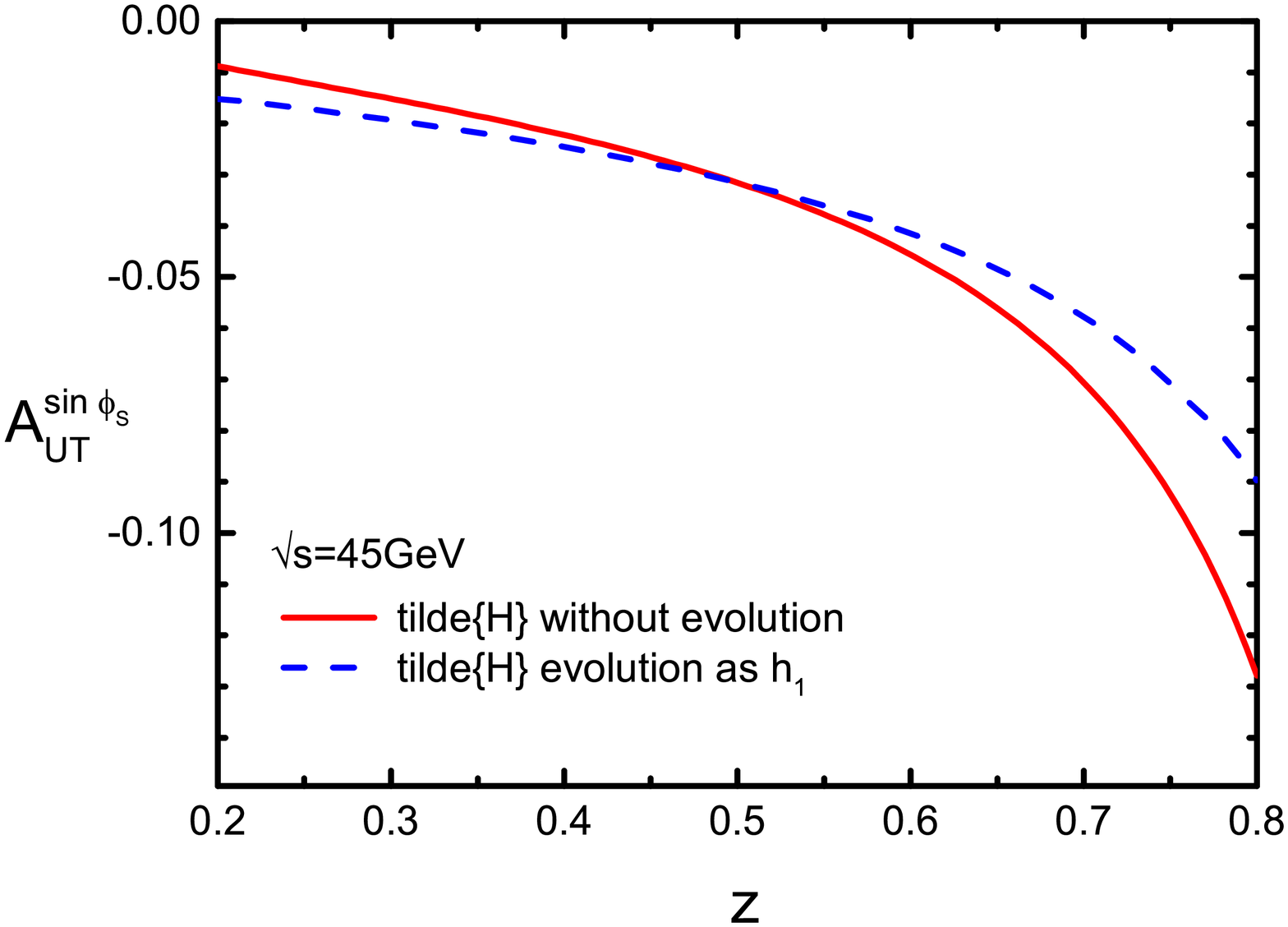}
  \caption{Transverse SSA $A_{UT}^{\sin\phi_S}$ of $\Lambda$ hyperon production in SIDIS at EIC for $\sqrt{s}=45$ GeV. The left and the right panels show the $x$-dependent and the z-dependent asymmetry, respectively.}
\label{fig:eic}
\end{figure}

\begin{figure}
  \centering
  \includegraphics[width=7.5cm]{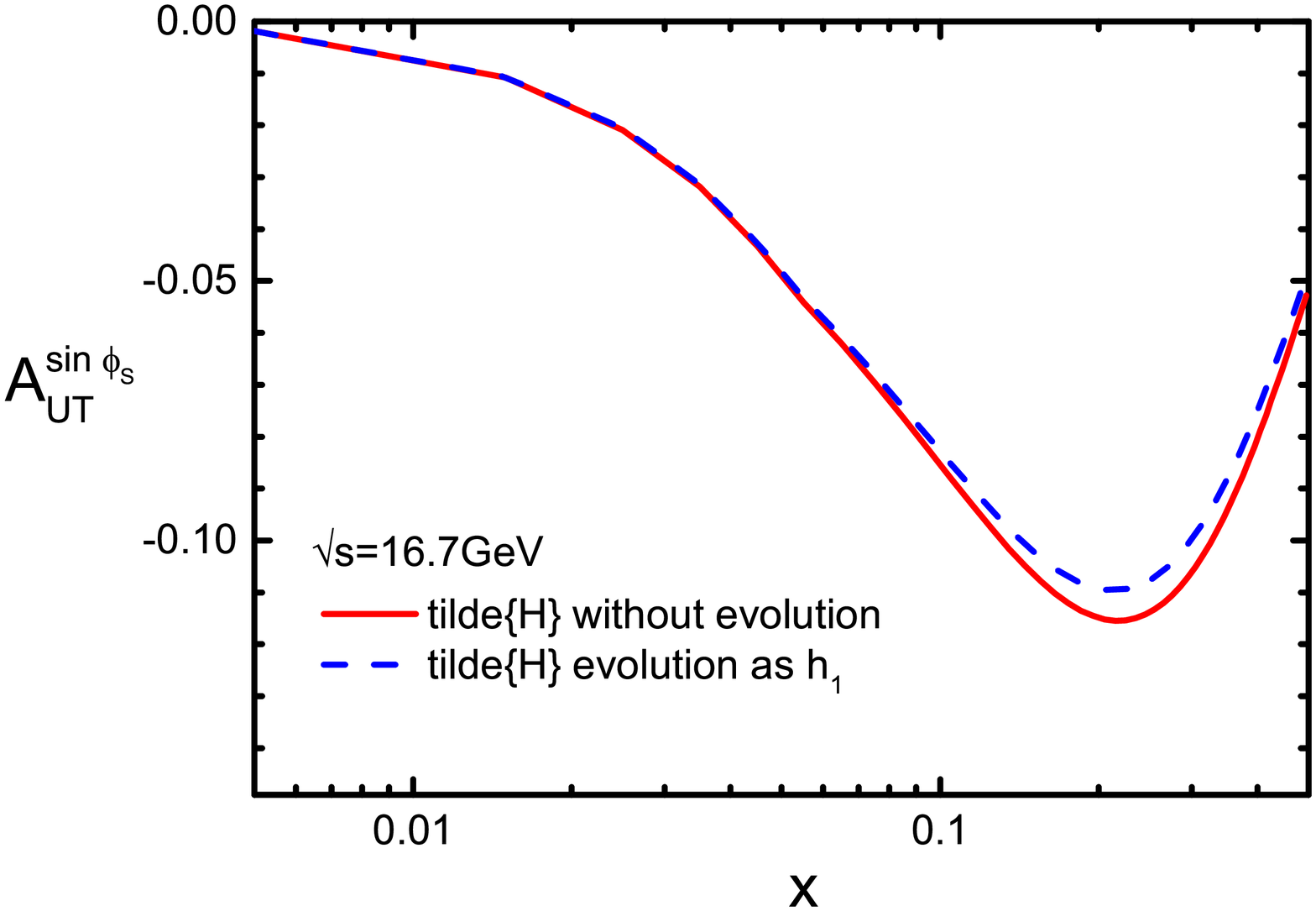}
  \includegraphics[width=7.5cm]{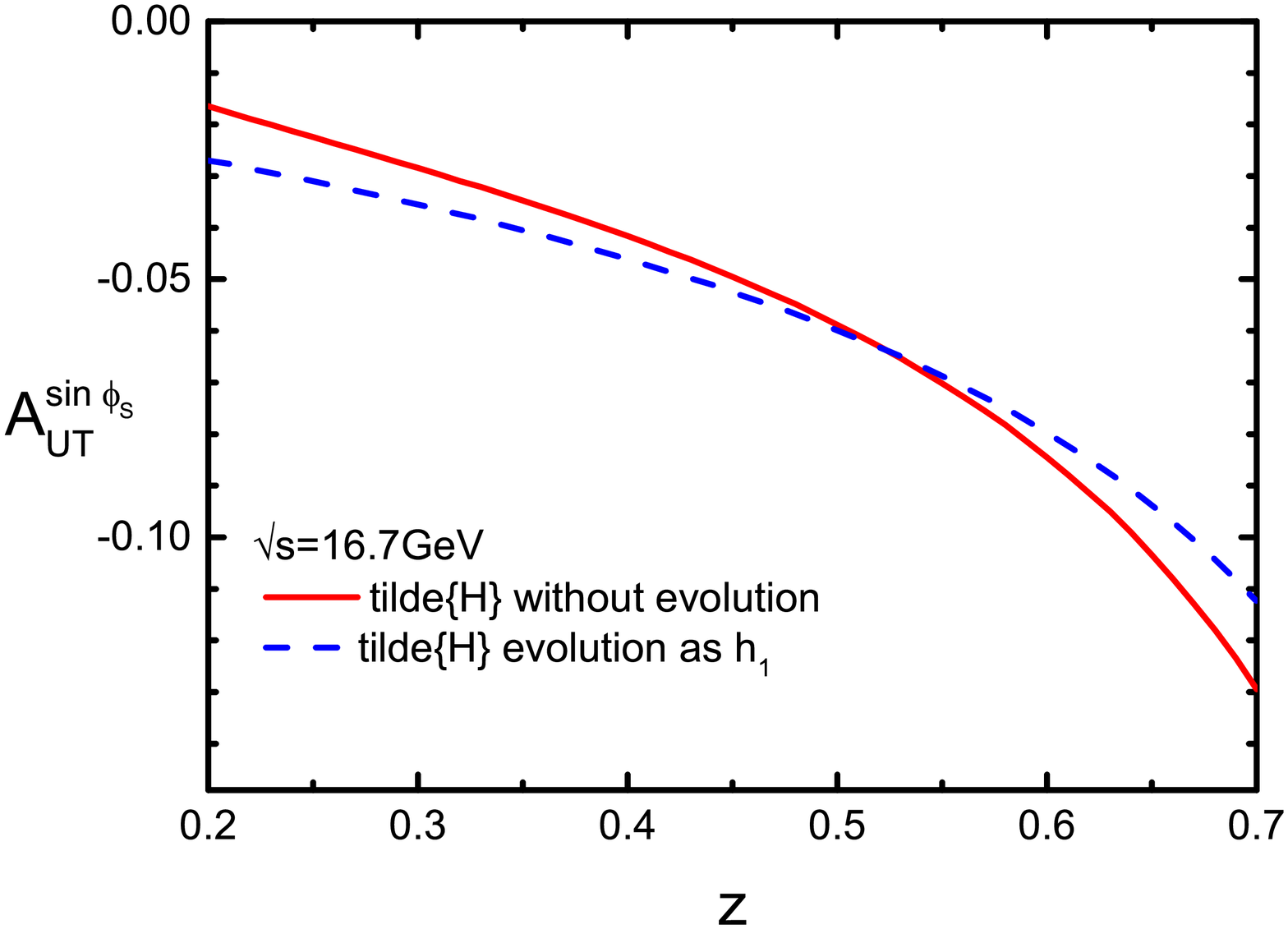}
  \caption{Transverse SSA $A_{UT}^{\sin\phi_S}$ of $\Lambda$ hyperon production in SIDIS at EicC for $\sqrt{s}=16.7$ GeV. The left and the right panels show the $x$-dependent and the z-dependent asymmetry, respectively.}
\label{fig:eicC}
\end{figure}

The numerical results of the $\sin \phi_S$ asymmetries for $\Lambda$ hyperon production at EIC and EicC are shown in Fig.~\ref{fig:eic} and Fig.~\ref{fig:eicC}, respectively.
The left panel and the right panel plot the asymmetries as functions of $x$ and $z$.
The dashed lines depict the results simultaneously evolving the fragmentation function $\tilde{H}$ and distribution function $h_1(x)$.
The solid lines denote the asymmetries without considering the evolution of the fragmentation functions $\tilde{H}$.
We find that the $\sin\phi_S$ asymmetries for the $\Lambda$ hyperon production are negative.
The magnitude of the asymmetry at EIC and EicC is around 0.1, which is quite sizable.
We also find that the $x$-dependent asymmetries have a peak at the intermediate $x$ region, around $x\approx 0.2$, and the magnitude of the $z$-dependent asymmetry increases with increasing $z$.
Compared our results with the the same $\sin \phi_S$ asymmetry in pion production~\cite{Wang:2016tix},
we find that the signs of asymmetry are both are negative;
However, the magnitude of the asymmetry in $\Lambda$ production is several times larger than the one in pion production.
The difference may be caused by the factor $M_h\over Q$ appearing in Eq.~(\ref{eq:futcol2}) as $M_\Lambda$ is almost one order of magnitude larger than $M_\pi$.
Therefore, there could be a great opportunity to measure the $\sin\phi_S$ asymmetry in $\Lambda$ production at a future EIC and EicC.

An other important observation is that the evolution effect for the $\sin\phi_S$ asymmetry is substantial in certain kinematical region both at EIC and EicC.
First of all, as shown by the dashed lines in Figs.~\ref{fig:eic} and~\ref{fig:eicC}, the magnitudes of the $x$-dependent and $z$-dependent asymmetries for $\Lambda$ hyperon has changed substantially by the QCD evolution effect.
Secondly, at small-$x$ region ($x < 0.04$ at EIC) and ($x < 0.07$ at EicC) the evolution does not affect the $x$-dependent asymmetries, while the evolution effect is sizable in the intermediate-$x$ region.
For the $z$-dependence asymmetries, the evolution effect may be observed in the all $z$ region, where the evolution effect of size is smaller than non-evolution result in the region $z>0.5$ at EIC and EicC, thereby it should not be neglected.
Nevertheless, the evolution almost does not change the signs and the shapes of the asymmetries.

\begin{figure}
  \centering
  \includegraphics[width=7.5cm]{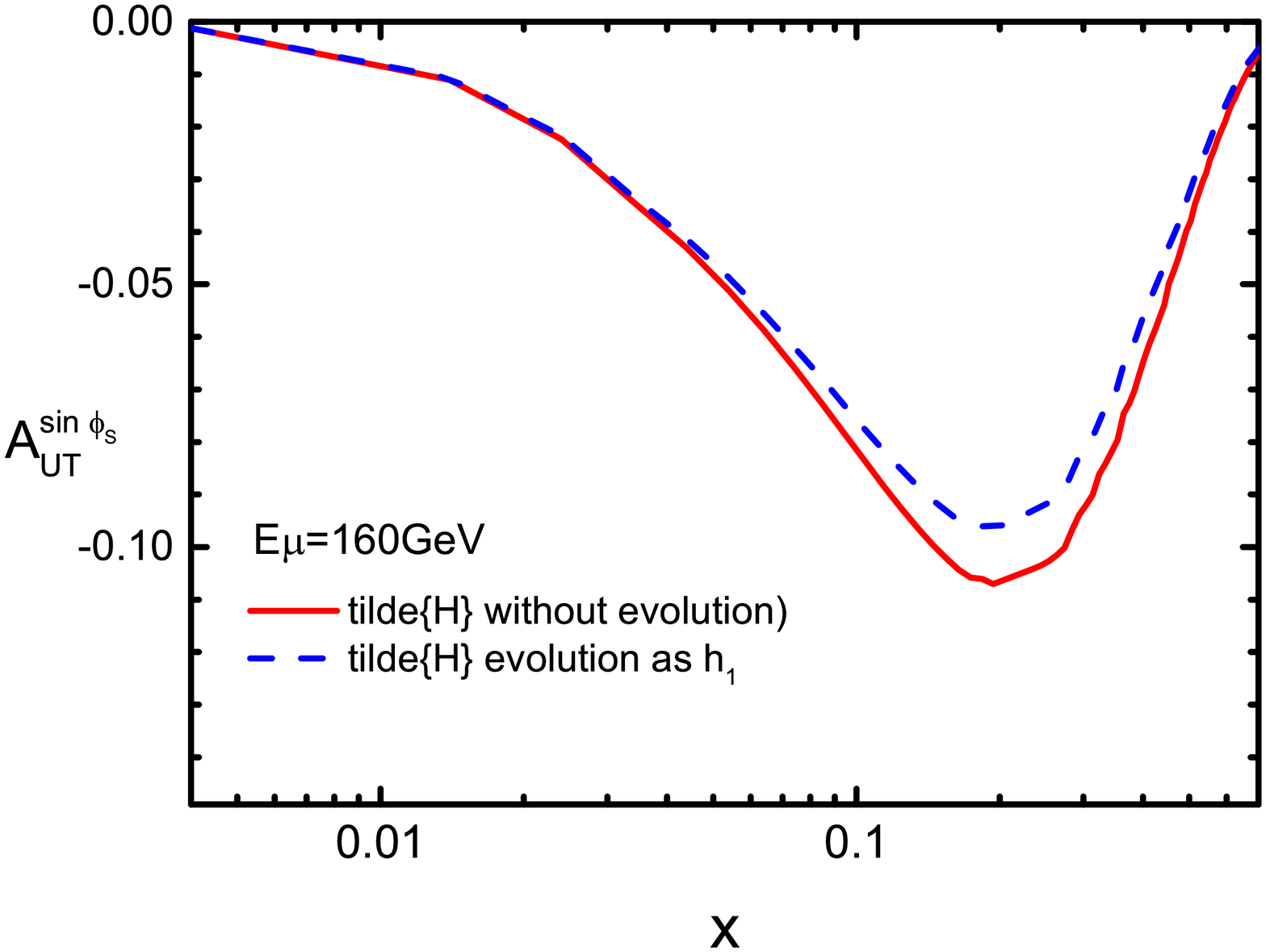}
  \includegraphics[width=7.5cm]{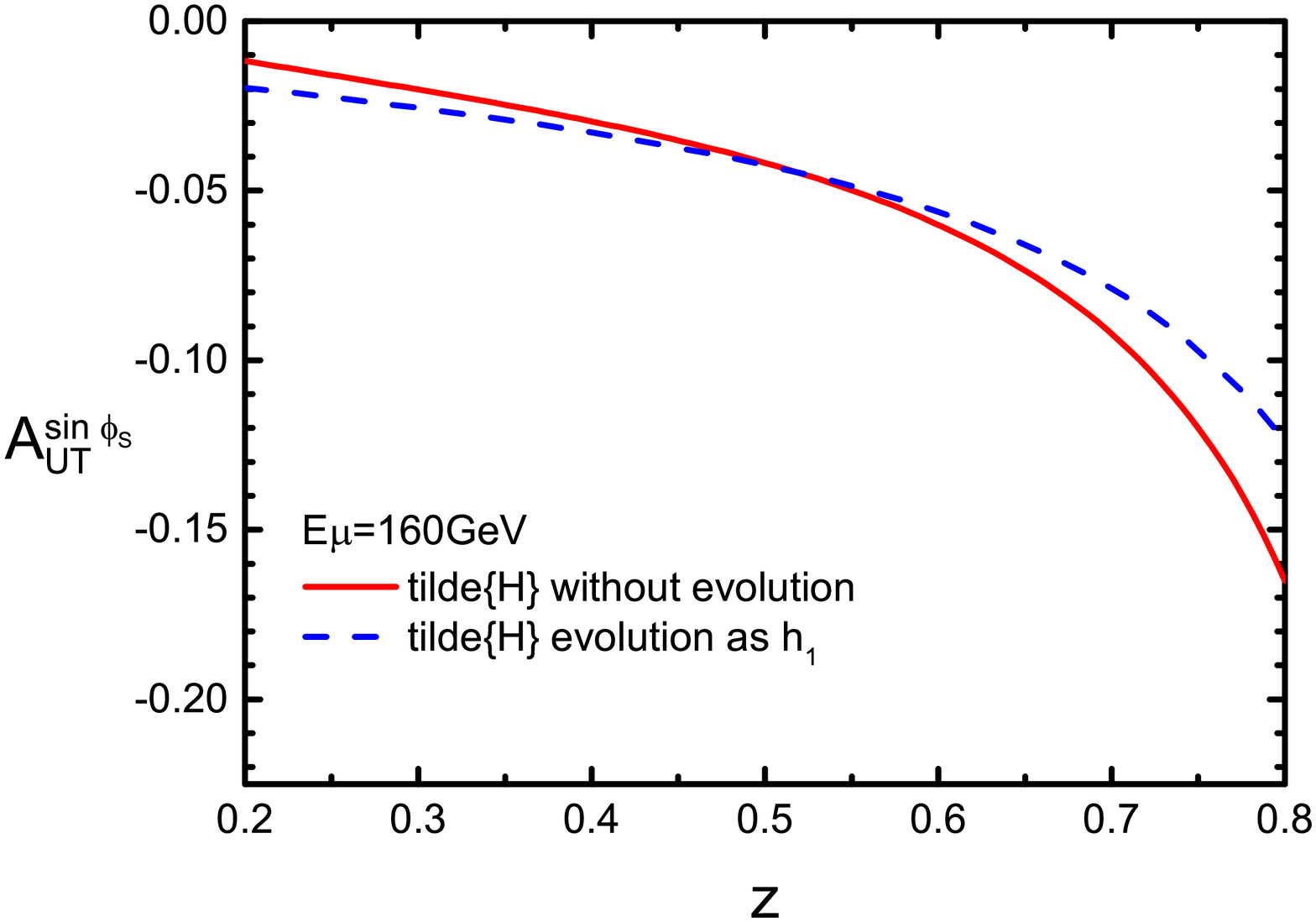}
  \caption{Transverse SSA $A_{UT}^{\sin\phi_S}$ of $\Lambda$ hyperon production in SIDIS at COMPASS for $E_\mu=160$ GeV. The left and the right panels show the $x$-dependent and the $z$-dependent asymmetry, respectively.}\label{fig:compass}
\end{figure}

Finally, we also estimate the transverse asymmetries for $\Lambda$ at COMPASS which applies a $160\mathrm{GeV}$ muon beam scattering off the nucleon target.
In this calculation, we adopt the following kinematical cults~\cite{Alekseev:2010rw}
\begin{align}
&Q^2>1 \mathrm{GeV}^2, \quad 0.004<x<0.7,\quad 0.1<y<0.9,\\\nonumber
&z > 0.2,\quad W>5\ \mathrm{GeV},\quad E_h > 1.5~\mathrm{GeV}\,.
\end{align}
The results of the $x$ and $z$-dependent asymmetries are depicted in the left panel and right panels in Fig.~\ref{fig:compass}, respectively.
The solid lines denote the asymmetries without considering the evolution of fragmentation functions $\tilde{H}$ in Eqs.~(\ref{eq:autx}) and (\ref{eq:autz}).
The dashed lines correspond to the evolution of $\tilde{H}^q(z)$ as $h_1$.
We find that the overall tendency of the asymmetries at the COMPASS are similar to that at EIC and EicC. The evolution effect for $x$-dependent asymmetry may be observed in the region $x>0.05$, and $z$-dependent asymmetry is larger than that of EIC and EicC.

Some comments are in order.
Firstly, in the model calculation we have calculated not only the $\tilde{H}^{\Lambda/q}$ for the up and down quarks, but also that for the strange quark, while in the phenomenological analysis of the $\sin\phi_S$ asymmetry we only have only considered the contributions from the up and down quark.
This is because currently there is no available information for the transversity of the sea quarks.
In several phenomenological studies~\cite{Lu:2011cw,Xue:2020xba}, the transversity of the sea quarks were included in the calculation through model assumptions, such as assuming the sea quark transversity is proportional to that of the valence quarks. In this work we refrain to do so since at this stage our result is a rough estimate of the asymmetry.
The $\tilde{H}^{\Lambda/s}$ could be measurable provided the transversity of the strange quark is sizable.
Secondly, in this work we consider the $\Lambda$ production in SIDIS process, which is normally more difficult to be measured than the meson production.
However, in this process it is not necessary to measure the polarization of the $\Lambda$.
Besides, as our estimate indicates, the asymmetry is quite sizable at EIC, EicC and COPMASS due to the factor $M_\Lambda/Q$ is much larger than $M_\pi/Q$.
Thus, we expect that the $\sin\phi_{S}$ asymmetry of $\Lambda$ production in SIDIS can be measured with the help of high statistic of the future $ep$ facilities.

\section{Conclusion}\label{conclusion}

In this work, we have studied the single-spin ${\sin\phi_{S}}$ asymmetry of the $\Lambda$ production in SIDIS off an transversely polarized proton target.
We have calculated the twist-3 T-odd quark-gluon-quark fragmentation function $\tilde{H} $ of the $\Lambda$ hyperon with two different types of diquark spectator model by considering both scalar and axial-vector diquarks~\cite{Jakob:1997wg,Bacchetta:2008af}.
The relation between the quark flavors and diquark types for the fragmentation functions, motivated by the SU(6) symmetric wave functions of the $\Lambda$ hyperon, has been taken into account to provide results for different flavors.
In addition, we have included the leading-order evolution effects for the fragmentation functions.
Using the numerical results of $\tilde H(z)$, we have estimated the SSA $A_{\mathrm{UT}}^{\sin\phi_{S}}$ in SIDIS at the kinematics of EIC, EicC and COMPASS.
Our calculation shows that the estimated $\sin\phi_{S}$ asymmetry for the $\Lambda$ production in SIDIS is sizable, around 0.1.
The sign of the asymmetry is negative in the entire $x$ and $z$ region.
We also find that the inclusion of the evolution effects of $\tilde{H}$ can change the shape and size of the asymmetry at the intermediate region of $x$ and $z$.
The evolution effects should be important for the interpretation of future experimental data.
In conclusion, the $\sin\phi_{S}$ asymmetries of $\Lambda$ production in SIDIS may be measured at the kinematics of  EIC, EicC and COMPASS, which provide a feasible way to  access the twist-3 collinear $\tilde{H}(z)$ of the $\Lambda$ hyperon via the $\sin\phi_S$ within the collinear framework.

\section*{Acknowledgements}
This work is partially supported by the Shandong Provincial Natural Science Foundation, China(Grants No. ZR2020QA081) and National Natural Science
Foundation of China (Grants No. 11575043,11905187,11847217). X. Wang is supported by the China Postdoctoral Science Foundation under Grant No.~2018M640680 and the Academic Improvement Project of Zhengzhou University.

\subsection{Appendix: Double $l$ integrals}
\begin{align}
\int{d^4l}{l^\mu\,l^\nu\,\delta(l^2)\delta((k-l)^2-m^2)\over ((k-P_h-l)^2-m_s^2)}=\mathcal{AA}k^\mu\,k^\nu+\mathcal{BB}P^\mu\,P^\nu+\mathcal{AB}(k^\mu\,P^\nu+P^\mu\,k^\nu)+g^{\mu\nu}\mathcal{W}_1
\end{align}
where
\begin{align*}
 \mathcal{AA}&=-{(k^2-m^2) [k^2 M^2 (3 \mathcal{A}+\mathcal{B})-2(\mathcal{B}-\mathcal{A}) (k\cdot P)^2-\mathcal{B} M^2 k\cdot P]\over k^2\lambda(m_s,M)}\\
 \mathcal{BB}&=-{(k^2-m^2)((\mathcal{A}+3 \mathcal{B})k^2-2 \mathcal{B} k\cdot P)\over \lambda(m_s,M)}\\
 \mathcal{AB}&=-{(k^2-m^2)(2 \mathcal{B} M^2-(\mathcal{A}+3 \mathcal{B})k\cdot P)\over \lambda(m_s,M)}\\
 \mathcal{W}_1&=-{(\mathcal{A}+\mathcal{B}) (k^2-m^2)\over 4}
\end{align*}
Note that $4k^2M^2-4(k\cdot P)^2=(4k^2M^2 - (k^2 +M^2 - m_s^2)^2)=-\lambda(m_s,M)$.
\begin{align}
\int{d^4l}&{l^\mu\,l^\nu\,\delta(l^2)\delta((k-l)^2-m^2)\over ((k-P_h-l)^2-m_s^2)(-l\cdot n_++i\epsilon)}=\mathcal{CC}_fk^\mu\,k^\nu+\mathcal{DD}_fP^\mu\,P^\nu+\mathcal{EE}_fn_+^\mu\,n_+^\nu\\
&+\mathcal{CD}_f(k^\mu\,P^\nu+P^\mu\,k^\nu)+\mathcal{CE}_f(k^\mu\,n_+^\nu+n_+^\mu\,k^\nu)
+\mathcal{DE}_f(n_+^\mu\,P^\nu+P^\mu\,n_+^\nu)+g^{\mu\nu}\mathcal{W}_2
\end{align}
where
\begin{align}
 \mathcal{CC}_fk^-=&-{2z (\mathcal{A}- \mathcal{B}) k\cdot P-2\mathcal{A} k^2 z-\mathcal{A} M_\Lambda^2+\mathcal{B} M_\Lambda^2 z+\mathcal{C}k^- (z-1) z (k^2-m^2)\over z^2\bm{k}_T^2}\displaybreak[0]\\
 \mathcal{CD}_fk^-=&{\mathcal{A} k^2 z-2 \mathcal{A} k^2-2 \mathcal{B} k\cdot P+ \mathcal{B} M_\Lambda^2+\mathcal{C}k^- (z-1) (k^2-m^2)\over z^2\bm{k}_T^2}\displaybreak[0]\\
 \mathcal{CE}_fk^-k^-=&-{2 (k\cdot P (\mathcal{B} (k^2 z^2+ M_\Lambda^2 (z-1))-\mathcal{A} k^2 z^2)+k^2 (\mathcal{A} (k^2 z^2+M_\Lambda^2 (z-1))-\mathcal{B} M_\Lambda^2 z^2))\over 2z^3\bm{k}_T^2}\displaybreak[0]\\
 &-{\mathcal{C} k^-(k^2-m^2) (k^2 z^2-2 z^2 k\cdot P+M_\Lambda^2 (2 z-1))\over 2z^3\bm{k}_T^2}\displaybreak[0]
\end{align}

\end{document}